\documentclass{iaus}
\usepackage{graphicx}
\usepackage{color}

\title[Star cluster formation in colliding galaxies] %% give here short title %%
{Shock-induced star cluster formation in colliding galaxies}

\author[Saitoh et al.]{Takayuki \textsc{R.Saitoh}$^{1}$,
        Hiroshi \textsc{Daisaka}$^2$,
        Eiichiro \textsc{Kokubo}$^{1,3,4}$,
        Junichiro \textsc{Makino}$^{1,3,4}$,
        Takashi \textsc{Okamoto}$^5$,
        Kohji \textsc{Tomisaka}$^{1,3,4}$,
        Keiichi \textsc{Wada}$^{3,6}$, \and 
        Naoki \textsc{Yoshida}$^7$
}

\affiliation{
{$^1$ Division of Theoretical Astronomy, National Astronomical
Observatory of Japan, 2--21--1 Osawa, Mitaka-shi, Tokyo 181--8588.} \\
{$^2$ Graduate School of Commerce and Management, Hitotsubashi University, Naka
2--1 Kunitachi-shi, Tokyo 186--8601} \\
{$^3$ Center for Computational Astrophysics, National Astronomical Observatory
of Japan, 2--21--1 Osawa, Mitaka-shi, Tokyo 181--8588} \\
{$^4$ School of Physical Sciences, Graduate University of Advanced Study
(SOKENDAI), 2--21--1 Osawa, Mitaka-shi, Tokyo 181--8588} \\
{$^5$ Center for Computational Sciences, University of Tsukuba 1--1--1,Tennodai,
Tsukuba, Ibaraki 305--8577,Japan} \\
{$^6$ Graduate School of Science and Engineering, Kagoshima University,
1--21--30 Korimoto, Kagoshima, Kagoshima 890--8580.} \\
{$^7$ Institute for the Physics and Mathematics of the Universe, University of
Tokyo, 5--1--5 Kashiwanoha, Kashiwa, Chiba 277--8568, Japan}
\\email: {\tt saitoh.takayuki@nao.ac.jp}
}

\pubyear{2010}
\volume{270}  %% insert here IAU Symposium No.
\jname{Computational Star Formation}
\editors{J. Alves, B. G. Elmegreen, J. M. Girart, \& V. Trimble, eds.}
\begin{document}

\maketitle

\begin{abstract}
We studied the formation process of star clusters using high-resolution
$N$-body/smoothed particle hydrodynamcs simulations of colliding galaxies.  The
total number of particles is $1.2\times10^8$ for our high resolution run.  The
gravitational softening is $5~{\rm pc}$ and we allow gas to cool down to $\sim
10~{\rm K}$.  During the first encounter of the collision, a giant filament
consists of cold and dense gas found between the progenitors by shock
compression. A vigorous starburst took place in the filament, resulting in the
formation of star clusters.  The mass of these star clusters ranges
from $10^{5-8}~{M_{\odot}}$.  These star clusters formed hierarchically: at first
small star clusters formed, and then they merged via gravity, resulting in
larger star clusters.

\end{abstract}

\firstsection % if your document starts with a section,
\section{Introduction}

Merging galaxies contain many young star clusters (e.g., \cite[Whitmore
2003]{Whitmore2003}).  These star clusters could potentially evolve into the
present day metal-rich globular clusters and so they are widely accepted to be a
good candidate for globular cluster progenitors.  Galaxy-galaxy merger is
therefore considered as one of the formation channels of globular clusters
(\cite[Schweizer 1987]{Schweizer1987}).

There have been a number of numerical studies of merging galaxies. There have
been, however, only a few numerical studies of star cluster formation in merging
galaxies even though it has been shown that resolving a cloudy/multiphase
interstellar medium (ISM) and/or clustered star formation can have important
consequences for the formation history of early-type galaxies (\cite[e.g. Bois
et al.  2010]{Bois+2010}). Some of the existing studies adopted sub-grid models
of star cluster formation (\cite[e.g., Bekki \& Couch 2001; Li et al.
2004]{BekkiCouch2001, Li+2004}), while more recently their formation has been
captured directly (e.g.  using the sticky particle method in \cite[Baurnaud et
al.  2008]{Baurnaud+2008}). Here we report the result of merger simulations that
capture the multiphase nature of the ISM and include realistic models of star
formation and feedback.

\section{Method}

We prepared two identical progenitor galaxies and then let them merge from a
parabolic and coplanar configuration.  The mass of components in one progenitor
galaxy is $10^{11}~M_{\odot}$ for the dark matter halo, $4.7 \times
10^9~M_{\odot}$ for the stellar disk and $1.8 \times 10^9~M_{\odot}$ for the gas
disk.  The galaxies are modeled using both $N$-body and smoothed
particle hydrodynamics (SPH) particles.  We employed $1.2 \times 10^8$ particles
for the two progenitor galaxies for the finest runs where the corresponding mass
of each particle is $1.9 \times 10^3~M_{\odot}$ for both $N$-body and SPH
particles.

Numerical simulations were performed by our $N$-body/SPH code, {\tt ASURA}.  We
adopted the FAST scheme (\cite[Saitoh \& Makino 2010]{SaitohMakino2010}) that
accelerates the time integration of a self-gravitating fluid by asynchronously
integrating gravity and hydrodynamical interactions. In addition, we used a
time-step limiter for the time-integration of SPH particles with individual
time-steps which enforces the differences of time steps in neighboring SPH
particles to be small enough so that the SPH particles can correctly evolve
under strong shocks (\cite[Saitoh \& Makino 2009]{SaitohMakino2009}).

The high mass resolution allows us to adopt the ISM model with the wide
temperature range down to $10~{\rm K}$ and realistic conditions of star
formation ($\rho > 100~{\rm cm^{-3}}$, $T <100~{\rm K}$ and $\nabla \cdot v <
0$, where $\rho,T$ and $v$ indicate density, temperature and velocity,
respectively). The feedback from type II supernovae (SNe) was also taken into
account.  These models are the same as those in \cite[Saitoh et al.  (2008;
2009)]{Saitoh+2008,Saitoh+2009}.

\section{Results}

At the first encounter (after $\sim 420~{\rm Myr}$ from the beginning of the
simulations), strong shocks took place.  These shocks induced the formation of a
large filament of cold and dense gas between the two galaxies.  This filament
quickly cooled and fragmented to a number of small, high-density clumps, and a
large-scale starburst took place in the filament (see \cite[Saitoh et al.
2009]{Saitoh+2009}). This starburst did not take place in previous studies in
which gas cooling was inhibited below $10^4~{\rm K}$. This starburst continued
for several $10~{\rm Myr}$ and was then quenched by the SN feedback.  A number
of star clusters formed during this starburst.

\begin{figure}[htb]
\begin{center}
\includegraphics[width=5.0in]{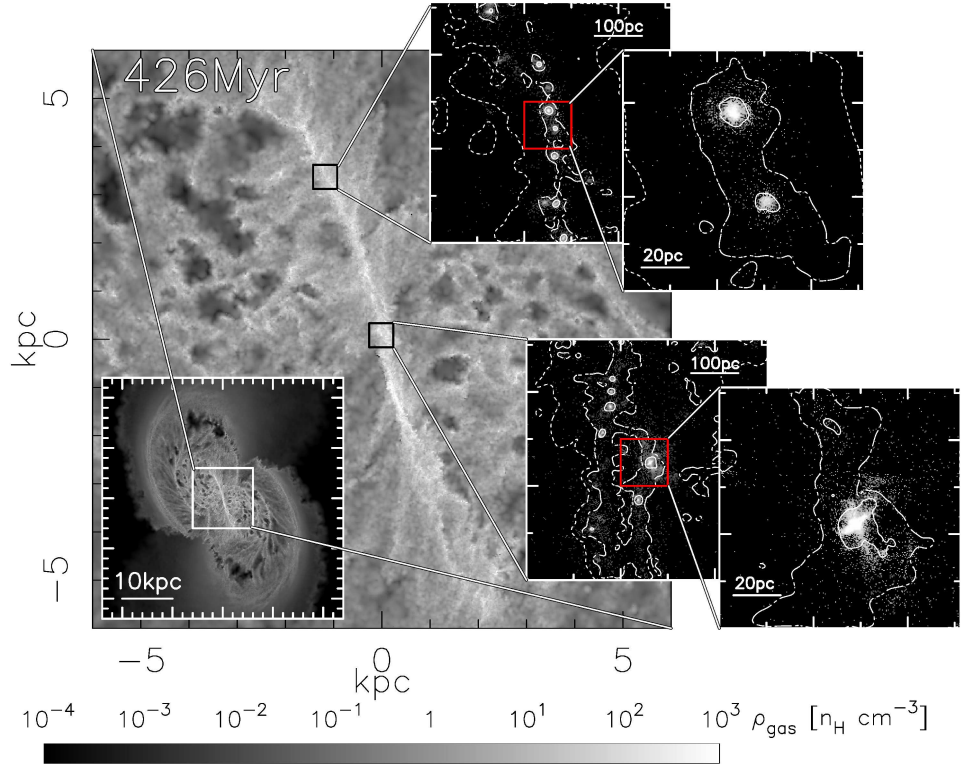} 
\caption{Distributions of gas and stars in the early phase of the star cluster
formation ($t = 426~{\rm Myr}$).  Main panel shows gas density map of the
mid-plane in $12~{\rm kpc}\times12~{\rm kpc}$ regions, while the bottom-left
inset displays four times larger scale than that of the main panel. Closeup
views of star-cluster forming regions show surface stellar density maps with
contours of surface gas density in $500~{\rm pc}\times500~{\rm pc}$ and
$100~{\rm pc}\times100~{\rm pc}$.
}
\label{Saitoh_Fig1}
\end{center}
\end{figure}

\begin{figure}
\begin{center}
\includegraphics[width=2.8in]{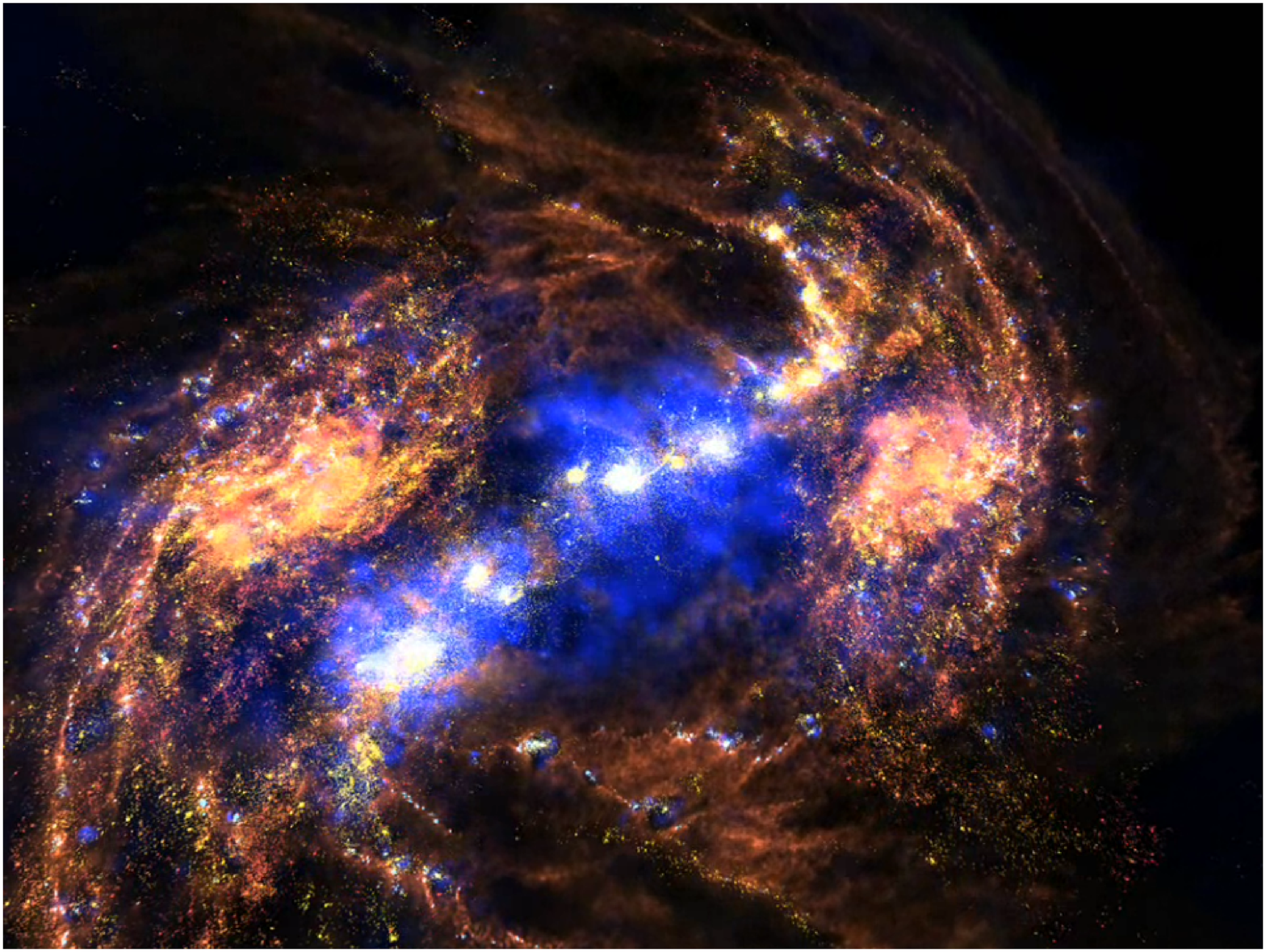} 
\caption{Distributions of star clusters formed during the first encounter at $t \sim
460~{\rm Myr}$.
Blown and blue means cold and hot gas phases, whereas white points are young
stars.  The visualization was done by Takaaki Takeda. 
}
\label{Saitoh_Fig2}
\end{center}
\end{figure}

Figure \ref{Saitoh_Fig1} shows the snapshots of the early phase of the star cluster
formation where we can see that a number of small star clusters formed along the
gas filament. The late phase of the star cluster formation ($t \geq 440~{\rm
Myr}$) was mainly driven by mergers of star clusters without gas, since gas in
the filament was blown out by SNe.  Hence star clusters grew hierarchically
through gravitational mergers of smaller star clusters. There is no star cluster
with the mass $\ge 10^7~{\rm M_{\odot}}$ which forms from an instantaneous
collapse of a single large cloud. At the final phase (after $40~{\rm Myr}$ from
the start of the starburst), we found several tens of star clusters between two
progenitor galaxies (see figure \ref{Saitoh_Fig2}).  This is because they were
formed only from the gas in the shock-induced filament.

\begin{figure}
\begin{center}
\includegraphics[width=2.8in]{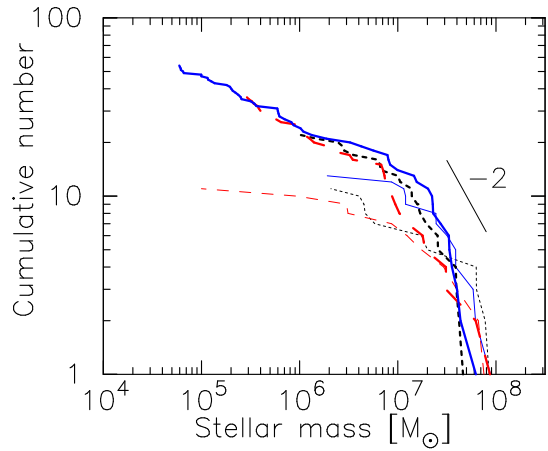} 
\caption{Cumulative mass functions of star cluster systems at $t = 460~{\rm Myr}$.
Thick and thin curves indicate cumulative mass functions of star clusters in runs with
$\epsilon = 5~{\rm pc}$ and $20~{\rm pc}$, respectively.  Solid, dashed, dotted
curves represent high, middle, low mass resolution runs, respectively.
}
\label{Saitoh_Fig3}
\end{center}
\end{figure}

Figure \ref{Saitoh_Fig3} shows cumulative mass functions of star cluster systems. The
mass of star clusters ranges from $10^{5-8}~{\rm M_{\odot}}$.  The mass function for
masses larger than $10^7~{\rm M_{\odot}}$ is almost independent of mass and
spatial resolutions.  The slope of this high mass region is close to a power
law function with an index of -2, which is consistent with observations
(e.g., \cite[Whitmore et al.  1999]{Whitmore+1999}). Since the formation of star
clusters is driven by mergers, the shape of the mass function becomes a power law
reflecting the scale free nature of gravity (\cite[Elmegreen 2006]{Elmegreen2006}).

\section{Summary}
We performed high resolution simulations of merging galaxies that capture the
multiphase ISM and the realistic modelings of star formation and of SN feedback.
In the simulations, we found that a number of star clusters formed during the
starburst at the first encounter.  These star clusters grew via hierarchical
mergers.

\section*{Acknowledgements.}
Numerical simulations were carried out on Cray XT4 at CfCA of NAOJ. TRS is
financially supported by a Research Fellowship from the JSPS for Young Scientists.

\end{document}